\documentclass[11pt,a4paper]{article}
\usepackage{jcappub}
\usepackage{amsmath,amssymb,epsf}
\usepackage{graphicx,color}

\def\bel#1{\begin{equation} \label{#1}}

\def\mc{\mathcal}
\def\be{\begin{equation}}
\def\ee{\end{equation}}
\def\bea{\begin{eqnarray}}
\def\eea{\end{eqnarray}}

\def\ltap{\ \raise.3ex\hbox{$<$\kern-.75em\lower1ex\hbox{$\sim$}}\ }
\def\gtap{\ \raise.3ex\hbox{$>$\kern-.75em\lower1ex\hbox{$\sim$}}\ }
\def\gl{\ \raise.5ex\hbox{$>$}\kern-.8em\lower.5ex\hbox{$<$}\ }
\def\roughly#1{\raise.3ex\hbox{$#1$\kern-.75em\lower1ex\hbox{$\sim$}}}

\def\hf{\frac12}

\def\cv{{{\cal{V}}}}

\newcommand\vo{\mathcal{V}}
\newcommand{\comments}[1]{}

\newcounter{oldcounter}
\addtocounter{equation}{1}
\title{$N$-flation with Hierarchically Light Axions in String Compactifications}
\author[a,b,c]{Michele Cicoli,}
\author[d]{Koushik Dutta,}
\author[e]{Anshuman Maharana}
\affiliation[a]{Dipartimento di Fisica ed Astronomia, Universit\`a di Bologna, \\ via Irnerio 46, 40126 Bologna, Italy}
\affiliation[b]{INFN, Sezione di Bologna, Italy}
\affiliation[c]{Abdus Salam ICTP, Strada Costiera 11, Trieste 34014, Italy}
\affiliation[d]{Theory Division, Saha Institute of Nuclear Physics, 1/AF Salt Lake, Kolkata, 700064, India }
\affiliation[e]{Harish Chandra Research Intitute, Chattnag Road, Jhunsi, Allahabad, 211019, India.}
\emailAdd{mcicoli@ictp.it}
\emailAdd{koushik.dutta@saha.ac.in}
\emailAdd{anshumanmaharana@hri.res.in}

\abstract{We propose a possible embedding of axionic $N$-flation in type IIB string compactifications
where most of the K\"ahler moduli are stabilised by perturbative effects, and so
are hierarchically heavier than the corresponding $N\gg 1$ axions whose collective dynamics drives inflation.
This is achieved in the framework of the LARGE Volume Scenario for moduli stabilisation.
Our set-up can be used to realise a model of either large field inflation or quintessence, just by
varying the volume of the internal space which controls the scale of the axionic potential.
Both cases predict a very high scale of supersymmetry breaking.
A fully explicit stringy embedding of $N$-flation would require control over dangerous back-reaction effects due to a large number of species.
A viable reheating of the Standard Model degrees of freedom
can be achieved after the end of inflation due to the perturbative decay of the $N$ light axions which drive inflation.}

\begin{document}

\maketitle

\section{Introduction}

There has been immense progress in observational cosmology in recent years. The observations have put up a large number
of challenges for theorists. Measurements of the cosmic microwave background
are giving ever surmounting evidence that the early universe went
through an inflationary phase \cite{Planck}. Apart from providing a predictive theory for
density perturbations, inflation also gives appealing explanations
for many of the puzzles in the standard model of cosmology. In the simplest form of the inflationary paradigm,
the energy density of the universe is dominated by the potential energy of a scalar field $\varphi$ rolling along its potential.
Observations require the potential to be extremely flat. More quantitatively, one requires
that the inflaton potential $V(\varphi)$ satisfies the so-called `slow roll' conditions:\footnote{For a comprehensive review on inflation see \cite{Baumann:2009ds}.}
\be
\epsilon \equiv  \hf  M_P^2 \bigg( { V'(\varphi) \over V(\varphi) } \bigg)^2 \ll 1, \qquad
\eta \equiv M_P^2 \bigg( { V''(\varphi) \over V(\varphi) } \bigg) \ll 1\,.
\label{slowroll}
\ee

From the theoretical point of view, achieving a flat potential is a real challenge. In fact,
the equation defining $\eta$ in (\ref{slowroll}) can be written
in terms of the inflaton mass and the Hubble constant $H$ as:
\be
\label{mphi}
\eta = \frac{m^2_{\varphi}}{3 H^2} \ll 1\,.
\ee
However, similarly to the Higgs, it is hard to protect the inflaton mass against
large corrections of order $H$ which naturally emerge when integrating out any kind of ultraviolet (UV) physics.
Because of this UV sensitivity, inflationary model building can be trusted only within a
UV complete theory like string theory where the slow-roll parameters can be computed reliably.

Moreover, string theory gives rise to a plethora of scalar fields in four dimensions,
naturally providing a large number of inflaton-like fields (see \cite{micireview} for a recent review). These moduli fields also come along
with approximate symmetries which suppress higher dimensional operators. Two main examples are axions which enjoy
shift symmetries that are broken only by non-perturbative effects \cite{axionReview}, and K\"ahler moduli other than the
volume of the internal space since they effectively do not appear in the expression of the tree-level K\"ahler potential \cite{Cicoli:2011zz}.

Another important issue to trust inflationary model building in a multi-dimensional field space,
is the absence of any steep direction orthogonal to the inflaton trajectory which could ruin the inflationary dynamics.
This is achieved if all the moduli other than the inflaton develop a mass larger than the Hubble scale $H$:
\be
\label{mmod}
m^2_{\rm mod} \gg H^2 \gg m^2_{\varphi}\,.
\ee
Therefore full moduli stabilisation is a crucial ingredient in order to build concrete models of inflation in string theory.

A promising inflaton candidate is a single axion field
because it enjoys a perturbative shift symmetry \cite{Freese:1990rb}. However, its non-perturbatively generated potential
is flat enough to drive inflation only if its decay constant $f_a$ is larger than $M_P$ \cite{Planck,axionReview}.
Due to the impossibility to find such a large $f_a$ in the low-energy effective field theory of string
compactifications \cite{svr,svrw,banks,nima,StringAxions}, a lot of attention has been given to models with $N\gg 1$ light axions where
each axion has a sub-Planckian decay constant but the collective motion of all of them can effectively be trans-Planckian \cite{nfla}.

Despite axionic $N$-flation has been argued to be string-inspired due to the need of a large number of axions,
this inflationary model has not yet been successfully embedded in an explicit string compactification due to the
difficulty to create a hierarchy of scales between the axions and the K\"ahler moduli.\footnote{See however \cite{kallosh,grimm}
for different proposals to embed $N$-flation in string theory.}
In this paper, we propose a solution to this problem by presenting a model of $N$-flation
in type IIB flux compactifications where all the K\"ahler moduli associated to the axions which drive inflation,
are fixed by perturbative effects. At this level of approximation, due to their shift symmetry,
the axions are exactly massless, whereas the K\"ahler moduli develop non-zero masses which are exponentially
larger than the axionic masses generated by subleading non-perturbative effects.
Our construction is based on the LARGE Volume Scenario (LVS) \cite{LVS} where all the K\"ahler moduli
controlling the size of large internal four-cycles are stabilised by perturbative effects.

We shall show that this set-up can be used to obtain a model of either large field inflation or quintessence.
In the first case, the volume of the extra dimensions $\vo$ is of $\mc{O}(10^2)$ in string units,
whereas in the second case $\vo$ has to be increased to values of $\mc{O}(10^3)$ in string units.
Note that such a low value of $\vo$ implies a gravitino mass of order the scale of grand unification,
resulting in the impossibility to obtain low-energy supersymmetry. A possible way to obtain TeV-scale
supersymmetry would be to tune the tree-level superpotential by an appropriate choice of flux quanta.
However, this choice would also ruin the mass hierarchy between the axions and the K\"ahler moduli,
and so we shall not consider it. Thus we conclude that in our set-up axionic $N$-flation with heavy K\"ahler moduli
predicts high-scale supersymmetry breaking.

A problem associated with our construction, which is intrinsic to $N$-flation, is the fact that $N$ has to be very large. 
In our case, the requirement of generating $60$ e-foldings of inflation with the right amplitude of the density perturbations, 
fixes $N\sim \mc{O}(10^5)$ which for $\vo\sim\mc{O}(10^2)$ gives a huge density of cycles per unit volume. 
This may result in dangerous backreaction issues due to this large number of species.\footnote{Note that there is also a generic tension between large field inflationary models and dS entropy bounds \cite{EntropyBounds}.}
Therefore a fully explicit stringy embedding of $N$-flation following our moduli stabilisation mechanism should take these corrections properly into account.
Following \cite{nfla}, we also briefly comment on the fact that a large number of species renormalises the Planck mass.
The validity of the low-energy effective theory and topological properties of the extra dimensions can help to alleviate this
problem in string theory, even if a full solution would require a complete understanding
of all possible corrections in string compactifications.

We finally comment on the way the inflaton energy density gets transferred to Standard Model particles.
The details of this reheating process depend on the strength of the coupling of the inflaton to both visible and hidden
degrees of freedom which, in turn, depend on the localisation of the Standard Model within the internal manifold.
We shall argue that a viable reheating of the visible sector can occur if the Standard Models lives on branes
wrapping large cycles which control the size of the internal manifolds and are associated to the axions which drive inflation.

This paper is organised as follows. In section \ref{Sec2} we give a brief review of axionic $N$-flation.
Section \ref{Sec3} provides a general discussion of how axions acquire
masses in type IIB Calabi-Yau (CY) flux compactifications and motivates our model which is presented in section \ref{Sec4}.
The general formalism in this section should be helpful
in understanding aspects of LVS when the number of K\"ahler moduli is large.
Section \ref{Sec5} gives an analysis of various phenomenological
aspects of the model including a possible mechanism for reheating. We also explore the
possibility of using our set-up to drive quintessence. Finally, we conclude with
a discussion of our results.

\section{Basics of $N$-flation}
\label{Sec2}

\subsection{Inflation from a single axion}

Axions enjoy a shift symmetry which remains unbroken in perturbation theory.
This crucial property of axionic fields guarantees the absence of dangerous higher dimensional
Planck-suppressed operators which would generate corrections to the slow-roll parameter $\eta$
of order unity. Moreover, any kind of correction to the potential from low-energy loops must be proportional
to the non-perturbative effect that breaks the shift symmetry, and hence can
be expected to be small, even in the absence of supersymmetry. This makes axions appealing as
inflaton candidates \cite{axionReview}.

The potential for a single axion $\theta$ takes the form:
\be
\label{voneinst}
V(\theta) = \Lambda^4 \cos\left(\frac{\theta}{f}\right),
\ee
where the scale $\Lambda$ is dynamically generated by instanton effects:
\be
\label{instsc}
\Lambda^4 \propto M_P^4\,e^{-S_{\rm inst}},
\ee
$f$ is the axion decay constant and $\theta \in [0,2\pi f[$. For the potential (\ref{voneinst}) it is easy to see that the slow-roll parameters
\be
\label{slowf}
\epsilon, \   \eta \propto  \left(\frac{M_P}{f}\right)^2\,.
\ee
Thus to achieve slow-roll one requires $f \gtrsim M_P$. On the other hand, it
can be argued on fairly general grounds \cite{svr,svrw,banks,nima,StringAxions} that in string theory $f \ll M_P$.
Thus there seems to be a conflict between slow-roll inflation and the effective field theories accessible from string theory.
In fact, one could have imagined a no-go theorem for axion inflation in string theory.

\subsection{$N$-flation: simplest realisation and predictions}

The problem can be circumvented with the $N$-flation proposal \cite{nfla}, where
inflation is driven by the collective dynamics of a large number of axions
which is a natural outcome of string compactifications \cite{rajesh}.\footnote{See also \cite{Assisted,Mflation} for other models where the inflationary dynamics is determined by more than one field.}
In its simplest incarnation, the $N$-flation potential consists of
$N$ axions with equal decay constants and dynamically generated scales:
\be
V(\theta_i) = \sum_{i=1}^{N}   \Lambda^4 \cos\left(\frac{\theta_i}{f}\right)\,.
\ee
Each canonically normalised axion $\phi$ develops a mass $m = \Lambda^2/f$ around its minimum, and so the previous potential
can be rewritten as:
\be
V(\phi_i) = \frac{m^2}{2} \sum_{i=1}^{N} \phi_i^2= \frac{m^2}{2} \,\rho^2\,,
\label{Vphii}
\ee
where:
$$
\rho^2 = \sum_{i=1}^N \phi_i^2\,,
$$
is the radial field in polar coordinates. At the onset of inflation, the displacement of each axions from its minimum is of
order $f$, resulting in an effective shift for $\rho$ of order:
\be
\label{feef}
f_{\rm{eff}} \simeq \sqrt{N} f\,.
\ee
For large $N$ it is easily possible to achieve slow-roll. In fact, the slow-roll parameters become:
\be
\epsilon=\eta=\frac{2 M_P^2}{\rho^2}\simeq \frac{2 M_P^2}{f_{\rm{eff}}^2}\simeq \frac{2}{N} \left(\frac{M_P}{f}\right)^2\ll 1
\qquad\text{for}\qquad N\gg1\,.
\label{eps}
\ee
Inflation ends when $\epsilon$ and $\eta$ become of order unity, i.e. at $\rho\simeq\rho_{\rm end}=\sqrt{2}M_P$.
The number of e-foldings is given by:
\be
N_e=\frac{1}{M_P^2}\int_{\rho_{\rm end}}^\rho\frac{V}{V'}d\rho=\frac{\rho^2}{4 M_P^2}-\frac 12 \simeq \frac{1}{2\epsilon}\,.
\label{Ne}
\ee
Let us now turn to the predictions for the two main cosmological observables, the spectral index $n_s$ and the tensor-to-scalar ratio $r$:\footnote{The predictions of the simplest version of $N$-flation are similar to the ones of single field chaotic inflation with a simple mass term since the effective potential (\ref{Vphii}) close to the minimum is nothing but the sum of $N$ polynomial `chaotic' potentials.}
\be
n_s=1+2\eta-6\epsilon=1-4\epsilon\simeq 1-\frac{2}{N_e}
\qquad\text{and}\qquad r=16\epsilon\simeq \frac{8}{N_e}\,.
\ee
Requiring $60$ e-foldings of inflation, one obtains from (\ref{Ne}) a constraint on the number of axions:
\be
N\simeq 240 \left(\frac{M_P}{f}\right)^2\,,
\label{Nnumber}
\ee
while the prediction for $n_s$ and $r$ become: $n_s\simeq 0.967$ and $r\simeq 0.133$.
The corresponding values estimated from Planck + WMAP data are: $n_s=0.9624\pm 0.0075$ at $1\sigma$
and $r<0.12$ at $2\sigma$ \cite{Planck} showing a very good agreement with the predicted value of the
spectral index but a slight tension with the upper bound on $r$. We stress that this $2\sigma$ discrepancy is not a
big effect and could instead be interpreted as the sign of a potential discovery of large gravitational waves produced during inflation. 
In fact, the BICEP2 collaboration has recently released their data which might point towards values of $r$ of order $0.1-0.2$ \cite{BICEP2}.

Due to this large value of $r$, the Hubble constant during inflation $H_{\rm inf}$ turns out to be rather high.
In fact, the amplitude of the scalar power spectrum can be written in term of $H_{\rm inf}$ as (for $N_e\simeq 60$):
\be
\mc{P}_{\rm scal}=\frac{H_{\rm inf}^2}{8\pi^2\epsilon M_P^2} = \frac{H_{\rm inf}^2 N_e}{4\pi^2 M_P^2}
\simeq 2.64\cdot 10^{-37}H_{\rm inf}^2\,{\rm GeV}\,.
\ee
Requiring this expression to match the observed value $\mc{P}_{\rm scal}\simeq 2.7 \cdot 10^{-7}$,
one finds:
\be
H_{\rm inf}\simeq 1\cdot 10^{15}\,{\rm GeV}\,.
\ee
The Hubble constant can also be rewritten in term of $m$ as:
\be
H_{\rm inf}^2=\frac{V}{3 M_P^2}=\frac{m^2 \rho^2}{6 M_P^2}=\frac{N m^2}{6}\left(\frac{f}{M_P}\right)^2\,.
\label{H}
\ee
Plugging (\ref{Nnumber}) in (\ref{H}), we end up with a constraint on the mass parameter $m$:
\be
H_{\rm inf}^2\simeq 40 \,m^2\qquad\Rightarrow\qquad m\simeq 1.6 \cdot 10^{14}\,{\rm GeV}\,.
\label{mFix}
\ee

We point out that this model is robust when one relaxes the above approximations by considering a random distribution for the
dynamical scales and decay constants, and allowing for kinetic mixing of the axions via
a random matrix \cite{random}. The net effect of this more general treatment is that the spectral index becomes slightly more red.
We will explore $N$-flationary predictions in context of our construction in section \ref{Sec5}.

\subsection{$N$-flation: main challenges for model building}

The two main challenges to realise this inflationary scenario are the following ones:
\begin{itemize}
\item \textit{Renormalisation of the Planck mass}: In the presence of $N$ light axions,
the Planck mass $M_P$ gets loop corrections of order \cite{nfla}:
\be
\delta M_P^2 \simeq c\,\frac{N}{16\pi^2} M_P^2\,,
\label{MPren}
\ee
where $c$ is a coefficient which can be computed exactly only in a given UV complete model.
Due to this renormalisation of the Planck mass, also the expression (\ref{eps}) for $\epsilon$ and $\eta$ gets modified to:
\be
\epsilon =\eta \simeq \frac{2}{N} \left(\frac{M_P^{\rm ren }}{f}\right)^2 \simeq
\left(\frac{2}{N} + \frac{c}{8\pi^2}\right)\left(\frac{M_P}{f}\right)^2\,,
\label{epsRen}
\ee
showing that, if the coefficient $c$ is of order unity, one cannot use the large number $N$ to
obtain small slow-roll parameters for $f\ll M_P$.

Note that this is just a low-energy field theory argument which does not take into account the details of UV complete models.
In fact, in string realisations of axionic $N$-flation, the exact prefactor $c$ of the loop-correction in (\ref{MPren})
can in principle be explicitly computed. The leading order correction to $M_P^2$ comes from the dimensional reduction of the
10D $\mc{O}(\alpha'^3)$ $R^4$-term and is proportional to the Euler number of
the CY three-fold, $\chi=2(h^{1,1}-h^{1,2})$ and its volume in string units $\vo$ \cite{BBHL}:
\be
c_{\alpha'}\simeq \frac{\chi}{N\,\vo}\,.
\ee
This number can easily be much smaller than unity either by focusing on large volume, $\vo\gg 1$,
which is a necessary condition to trust the supergravity effective field theory, or, if $N=h^{1,1}$,
by choosing appropriate internal manifolds with $\chi\ll h^{1,1}$.

On the other hand, the leading order string loop correction to the Planck mass is expected to be proportional to the
Kaluza-Klein scale $M_{KK}\simeq M_P/\vo^{2/3}$, resulting in a coefficient $c$ of order \cite{bhk,bhp,ccq}:
\be
c_{g_s}\simeq \frac{g_s^2}{N} \left(\frac{M_{KK}}{M_P}\right)^2 \simeq \frac{g_s^2}{N \vo^{4/3}}\,,
\ee
which is again much smaller than unity for $g_s\ll 1$ and $\vo\gg 1$.

These observations imply that the requirement of validity of the low-energy effective theory and
topological properties of the extra dimensions can help to alleviate this problem in string theory,
even if a full solution would require a complete understanding of all possible $\alpha'$ and $g_s$ corrections to $M_P^2$
in string compactifications.

\item \textit{Hierarchy between axions and K\"ahler moduli}: The axions of string compactifications
which preserve $\mc{N}=1$ supersymmetry come along with supersymmetric partners, the K\"ahler moduli,
which would develop the same mass as the axions if the moduli are stabilised in a supersymmetric way. In this case,
one would not be able to decouple the K\"ahler moduli from the inflationary dynamics.
More severely, even if supersymmetry is broken but the moduli are fixed by non-perturbative effects,
both the axions and the K\"ahler moduli would acquire the same mass.
On the other hand, if the K\"ahler moduli are stabilised by perturbative effects, the axions would remain massless at this
level of approximation because of their shift symmetry. The axions would then develop exponentially suppressed masses at the non-perturbative level,
so realising the required hierarchy of mass scales with respect to the K\"ahler moduli.
In section \ref{Sec3} we shall discuss this issue more in detail and in section \ref{Sec4} we shall present a model where this perturbative moduli stabilisation mechanism is realised, leading to $N$ light axions suitable for inflationary model building.
\end{itemize}

\section{Moduli stabilisation and axion masses}
\label{Sec3}

\subsection{Hierarchically light axions from string compactifications}

The arena for our construction will be type IIB compactifications on a CY orientifold with imaginary self-dual $G_3$ flux threading its three-cycles.
In these compactifications the closed string moduli consist of the dilaton $S=e^{-\phi} + {\rm i} \,C_0$, the complex structure moduli $U_\alpha$, $\alpha=1,...,h^{1,2}$ and the K\"ahler moduli $T_i$, $i=1,...,h^{1,1}$ with:\footnote{To simplify our discussion we will consider CY orientifolds with $h^{1,1}_- = 0$.}
\be
\label{kaldef}
T_i \equiv \tau_i + {\rm i}\, \theta_i =  \hf \int_{\Sigma_i}  J \wedge J + {\rm i} \int_{\Sigma_i}  C_4\,,
\ee
where $J$ is the CY K\"ahler form and $\Sigma_i$ a set of linearly independent four-cycles.
The real parts of the K\"ahler moduli correspond to volumes of four-cycles in Einstein frame
and are often referred to as the geometric moduli. On the other hand, the complex components of the $T$-moduli are axionic fields which
come along with shift symmetries inherited from the gauge symmetries of the Ramond-Ramond potentials:
\be
\theta_i \to \theta_i + {\rm constant}\,.
\ee
This shift symmetry is exact in string perturbation theory and gets broken only by non-perturbative effects,
implying that the axions remain exactly massless at perturbative level.
The number of these axions depends on the topology of the extra dimensions since it is given by $h^{1,1}$
which for an ordinary CY three-fold is rather large: $h^{1,1} \sim \mc{O}(100)$.

On top of closed string axions, one can have also open string axions arising as phases $\psi$ of
matter fields $C = |C|\,e^{{\rm i}\psi}$ which live on fractional D3-branes at singularities and are charged under an anomalous $U(1)$ factor.
The anomalous $U(1)$ gets a St\"uckelberg mass of order the string scale $M_s$ by eating up a closed string axion.
Thus the low-energy theory below $M_s$ has an effective global Peccei-Quinn $U(1)$ symmetry
which gets broken when $|C|$ develops a non-zero VEV by D-term stabilisation. The axion $\psi$
is massless at the perturbative level because of its shift symmetry:
\be
\psi\to\psi+ q\,,
\ee
where $q$ is the $U(1)$-charge of $C$. Again this symmetry is broken only by non-perturbative effects,
implying that open string axions, similarly to the closed string ones, are exactly massless at perturbative level.
The number of these open string axions is more model-dependent since it depends on the number of D3-branes
at singularities allowed by tadpole cancellation. We shall therefore not consider this option and focus on the model-independent
case of closed string axions. We just mention, however, that one can have also magnetised D7-branes
wrapping internal four-cycles but in this case it can be shown that the axions eaten up by the Green-Schwarz
mechanism for anomaly cancellation are the open instead of the closed string ones \cite{Cicoli:2013ana,DMDR}.

Contrary to the axion fields, the geometric moduli $\tau_i$ are not protected by any symmetry,
and so can develop a potential at both perturbative and non-perturbative level, leading to two different situations:
\begin{enumerate}
\item Perturbative effects (like $\alpha'$ and $g_s$ corrections)
are naturally larger than non-perturbative ones (like contributions from instantons or gaugino condensation on D7-branes).
In this case, the $\tau_i$ can be stabilised at leading perturbative level, while
the axions $\theta_i$ are lifted only by subleading non-perturbative effects. Given that the $\tau_i$
and the $\theta_i$ are fixed by different effects, their masses tend to be different with $m_{\tau_i}\gg m_{\theta_i}$.

\item Both $\tau_i$ and $\theta_i$ can develop a mass of the same order of magnitude, $m_{\tau_i}\simeq m_{\theta_i}$,
only if they are both fixed by the same effects, which for the axions can only be the non-perturbative ones.
This can happen only if the perturbative effects are made negligible by tuning some underlying parameters.
Notice that in supergravity these masses are generically of order the gravitino mass $m_{3/2}$.
\end{enumerate}
These observations imply that one can obtain a hierarchy in mass scales between the geometric moduli
and the axions only when some moduli are fixed perturbatively. This situation can be rather generic because of
the following considerations \cite{StringAxions,joeaxion}:
\begin{itemize}
\item \emph{Naturalness}: The $S$ and $U$-moduli are fixed by the background flux $G_3$
which generates a tree-level superpotential of the form \cite{gukov}:\footnote{From now on,
we will formulate the effective supergravity theory in units of $M_P$.}
\be
\label{gvw}
W_{\rm tree} = \int G_3 \wedge \Omega\,.
\ee
Note that $W_{\rm tree}$ does not depend on the K\"ahler moduli because of the axionic shift symmetry
and the fact that $W$ has to be holomorphic (hence can have no dependence on $\overline{T}_i$).
Combining this observation with the no-scale property of the tree-level K\"ahler potential for the $T$-moduli:
\be
K_{\rm tree}=-2\ln\vo = -2 \ln\left(\frac{1}{3!}\int J\wedge J\wedge J\right)\,,
\ee
the F-term scalar potential:
\be
V = e^K \left( K^{I \bar{J}} D_{I} W {D_{\bar{J}} \overline{W}}  - 3 |W|^2 \right),
\label{sugrapot}
\ee
is exactly flat at tree-level along the K\"ahler moduli directions.
After integrating out the $S$ and $U$-moduli, the relative strength of perturbative and
non-perturbative effects to fix the $T$-moduli is controlled by the VEV of the tree-level
superpotential $W_0=\langle W_{\rm tree}\rangle$ since from the structure of (\ref{sugrapot}) one finds:
\be
R\equiv\frac{V_{\rm pert}}{V_{\rm non-pert}}\simeq \frac{|W_0| K_{\rm pert}}{W_{\rm non-pert}}\,,
\label{R}
\ee
with:
\be
K_{\rm pert}=K_{\alpha'}+K_{g_s}\qquad\text{and}\qquad
W_{\rm non-pert} = \sum_i A_i \,e^{-a_i T_i}\,.
\label{supka}
\ee
Given that both $\alpha'$ and $g_s$ corrections to $K$ depend on inverse powers of the geometric moduli
while $W_{\rm non-pert}$ is exponentially suppressed with respect to the $\tau_i$, if $|W_0|$ takes natural
values of order unity, the ratio in (\ref{R}) becomes $R\gg 1$, resulting in a situation where non-perturbative effects
are naturally smaller than the perturbative ones. Hence they can be neglected at leading order where the geometric moduli
are fixed due to the effects which generate $K_{\rm pert}$.

\item \emph{Zero-mode conditions}: Even if $|W_0|$ is tuned to exponentially small values, $|W_0|\sim\mc{O}(W_{\rm non-pert})$,
so that the ratio in (\ref{R}) becomes $R\ll 1$, in order to fix all the K\"ahler moduli by non-perturbative effects, one should check that
indeed all the $T$-moduli appear in $W_{\rm non-pert}$. However, an instanton contributes to the superpotential only if
some restrictive fermionic zero mode conditions are satisfied (similar considerations apply also for gaugino condensation).
The simplest situation involves an $O(1)$ instanton which is realised as a Euclidean D3-brane with vanishing gauge flux wrapping a rigid divisor which is transversally invariant under the orientifold action \cite{InstReview}. In the presence of deformation zero-modes, chiral intersections, or a non-vanishing gauge flux,
the contribution of the instanton to the superpotential is not automatically guaranteed. Let us briefly comment on each of these cases:
\begin{enumerate}
\item \emph{Deformation zero-modes}: If the four-cycle $\Sigma_4$ wrapped by the instanton is not rigid,
$W_{\rm non-pert}$ can depend on the corresponding geometric modulus only if its deformation modes
are soaked up by suitable fluxes \cite{Luca1,Luca2}. These deformation modes are counted by $h^{1,0}(\Sigma_4)$
which gives the number of Wilson-line moduli and $h^{2,0}(\Sigma_4)$ which gives the number of transversal deformations
of $\Sigma_4$. An instanton definitely contributes to $W_{\rm non-pert}$ only if it is rigid, i.e. $h^{1,0}=h^{2,0}=0$.

\item \emph{Chiral intersections}: Charged zero-modes can arise at the intersection
between an instanton and a chiral D7-brane stack with non-zero gauge flux. In this case,
the prefactor $A$ of the non-perturbative superpotential
$W_{\rm non-pert} = A \,e^{- a T}$ will depend on charged matter fields $\phi_i$: $A \sim \prod_i \phi_i$.
If this D7-brane stack corresponds to the visible sector, the VEV of each $\phi_i$
has to vanish in order to preserve the visible gauge group, resulting in $A = 0$ \cite{blumenhagen}.
This tension sets a strong constraint: there has to be no chiral intersection between the visible sector
and any instantonic brane. This implies that it is rather hard to fix the cycle
supporting the visible sector via non-perturbative effects.

\item \emph{Non-vanishing gauge flux}: If the total gauge flux $\mc{F} = F - B$ on an instanton D3-brane
is different from zero, the instanton configuration would not be invariant under the orientifold,
and so no contribution to $W_{\rm non-pert}$ would be generated. However, a non-vanishing total flux $\mc{F}$ could
be generated as a consequence of the cancellation of Freed-Witten (FW) anomalies which induces a half-integer flux
$F$ on any \emph{non-spin} four-cycle \cite{FW}. In fact, one could cancel this flux by choosing $B=F$,
but this choice of $B$-field would induce a non-vanishing $\mc{F}$ on any divisor which intersects the original one \cite{BBGW,CKMW}.
Hence, due to FW anomaly cancellation, it is generically rather difficult to generate more than one non-perturbative effect
for a set of intersecting four-cycles.\footnote{A possible way-out could be to focus on rank-2 $U(1)$ instantons \cite{Inaki}.}
\end{enumerate}
\end{itemize}
A straightforward way to satisfy all these constraints is to consider non-perturbative effects only on del Pezzo divisors
since they are rigid and, in a proper basis, admit only self intersections \cite{CKM}. Thus one would automatically have no problems with
deformation zero modes and FW anomaly cancellation. Moreover, there would be no chiral intersections with the visible sector
if this is supported by different cycles. These are actually the only non-perturbative effects needed to achieve full moduli stabilisation
in LVS models \cite{LVS} where all the non del Pezzo divisors are fixed by perturbative effects,
and so the corresponding axions remain light \cite{GeneralLVS}. \footnote{In LVS models where the visible sector is supported by a del Pezzo divisor, this geometric modulus can be fixed by either D-terms of string loop effects \cite{CMV}.}
On the other hand, in KKLT scenarios \cite{kklt} all the moduli are stabilised by non-perturbative effects assuming
that $W_{\rm non-pert}$ depends on all the $T$-moduli. In this case, all the axions are as heavy as the geometric moduli.
Let us now briefly describe the simplest realisation of each of these two different moduli stabilisation mechanisms.

\subsection{KKLT scenario and heavy axions}

In its simplest realisation, the KKLT construction involves a CY with a single
K\"ahler modulus $T\equiv \tau + {\rm i} \theta$. The K\"ahler and superpotential governing
its dynamics take the form \cite{kklt}:
\be
K = - 2 \ln  \cv  = -3 \ln ( T + \overline{T} )
\qquad\text{and}\qquad
W = W_0 + A\, e^{-aT},
\label{supkklt}
\ee
where $\cv$ is the CY volume in string units and $|W_0|$ is fine-tuned to exponentially small values so that
perturbative corrections to $K$ can be neglected. For the above $W$ and $K$ the scalar potential becomes:
\be
\label{kkltpot}
V_{\rm KKLT} = { 1 \over 8 \tau^3 } \left[ |A|^2 e^{-2a \tau} \left( {4 \over 3} a^2 \tau^2 + 4 a \tau \right)
+ 4a \tau |A W_0| e^{-a \tau} \cos ( a \theta + \psi) \right],
\ee
where $\psi$ is defined by $A \overline{W}_0 =  |A W_0| e^{i \psi}$. The potential depends on the axion only through the last term.
The minimisation with respect to the axion is easily done by setting $\theta = (\pi- \psi) / a$ so that the cosine attains its minimum value.
Substituting the VEV of $\theta$ in (\ref{kkltpot}) one obtains the potential for $\tau$:
\be
\label{taupot}
V_{\rm KKLT} =  { a |A| e^{-a\tau} \over 2 \tau^2 } \left( {1 \over 3} a |A| \tau e^{-a\tau} + |A|e^{-a \tau} - |W_{0}| \right).
\ee
which leads to a stable vacuum where both $\theta$ and $\tau$ have masses of the same order of magnitude:
\be
\label{kkltmas}
m_\theta \simeq m_{\tau} \simeq m_{3/2}\simeq { |W_0|\over \cv}M_P.
\ee
This is essentially a consequence of the fact that both $\tau$ and $\theta$ acquire
masses due to the same ingredient in the effective action: the non-perturbative
effect on the cycle associated with $T$. The mass of the axions and geometric
moduli also remain at the same scale for generalisations of the KKLT construction with multiple K\"ahler moduli \cite{joeaxion},
and so they are not suited to be used as inflatons.

\subsection{LVS scenario and light axions}
\label{LVSstab}

The simplest realisation of LVS is for a CY orientifold with two K\"ahler moduli,
$T_b \equiv\tau_b +  {\rm i}\theta_b$ and $T_s\equiv\tau_s +  {\rm i} \theta_s$. The CY volume has the Swiss-cheese form \cite{LVS}:
$$
\cv = \tau_b^{3/2} - \tau_s^{3/2}\,,
$$
where the blow-up mode $\tau_s$ is a del Pezzo divisor suitable to support non-perturbative effects.
In this case, the VEV of the flux superpotential $|W_0|$ is naturally of order unity,
implying that perturbative corrections to $K$ play a crucial r\^ole for moduli stabilisation.
The K\"ahler potential with the leading $\alpha'$ correction looks like \cite{BBHL}:\footnote{String loop corrections to $K$ can be shown to give rise to subdominant effects \cite{bhk,bhp,ccq}.}
\be
K=-2\ln\left(\cv+\frac{\xi}{2 g_s^{3/2}}\right)\simeq -2\ln\vo-\frac{\xi}{g_s^{3/2}\vo},
\label{KLVS}
\ee
where ${\xi}$ is proportional to the CY Euler number $\chi$, while the superpotential reads:
\be
\label{Lvssup}
W = W_0 + A_s\,e^{-a_s T_s}\,.
\ee
The scalar potential (\ref{sugrapot}) for $K$ and $W$ as in (\ref{KLVS}) and (\ref{Lvssup}) is a rather complicated function of the moduli
but simplifies considerably in the limit $\tau_b \gg \tau_s$:
\be
V_{\rm LVS}= \lambda_1\,\frac{\sqrt{\tau_s}\,e^{-2a_s\tau_s}}{\vo}
+\lambda_2 \frac{|W_0|\tau_s\,e^{-a_s\tau_s}}{\vo^2} \cos ( a_s \theta_s + \psi_s)+ \lambda_3\frac{|W_0|^2}{\vo^3}\,,
\label{Vlvs}
\ee
where $\lambda_1= 8 a_s^2 |A_s|^2/3$, $\lambda_2=4 a_s |A_s|$, $\lambda_3= 3\xi/(4 g_s^{3/2})$
and $\psi_s$ is defined by $A_s \overline{W}_0 =  |A_s W_0| e^{i \psi_s}$.
The potential (\ref{Vlvs}) admits a global minimum with:
\be
\theta_s = \frac{\pi- \psi_s}{a_s},\qquad \tau_s \simeq \left(\frac{3\xi}{2}\right)^{2/3}\frac{1}{g_s},
\qquad \vo \simeq \frac{3\sqrt{\tau_s}|W_0|}{4 a_s |A_s|}\,e^{a_s \tau_s}\sim |W_0|\,e^{a_s/g_s}\,,
\label{globalMin}
\ee
where $\tau_b\simeq \vo^{2/3}$ is exponentially large in $\tau_s$, justifying the limit $\tau_b \gg  \tau_s$.
Both the axion $\theta_s$ and the blow-up mode $\tau_s$ are fixed by the same non-perturbative effect,
and so develop a mass of the same order of magnitude, as in the previous KKLT case:
\be
\label{ms}
m_{\theta_s} \simeq m_{\tau_s}  \simeq m_{3/2}\simeq{ |W_0| \over \cv}M_P\,.
\ee
On the other hand, the large modulus $\tau_b$ is fixed by balancing the leading order $\alpha'$ correction to $K$ with the $T_s$-dependent
non-perturbative effect. Hence we expect this modulus to be heavier than the corresponding axion
which can develop a potential only due to non-perturbative effects in $T_b$. In fact, $\tau_b$ acquires a mass of order:
\be
\label{mb}
m_{\tau_b} \simeq { |W_0| \over \cv^{3/2}}M_P\,,
\ee
while the potential (\ref{Vlvs}) does not depend on $\theta_b$, implying that this axion is exactly massless at this level of approximation.
By including a subleading $T_b$-dependent non-perturbative superpotential of the form $W_{\rm sub} = A_b\,e^{-a_b T_b}$,
also this remaining flat direction is lifted, giving to $\theta_b$ a tiny mass of order:
\be
\label{mbaxion}
m_{\theta_b} \simeq  \sqrt{|W_0|} M_P\,e^{-\frac{a_b}{2} \tau_b} \simeq \sqrt{|W_0|} M_P\,e^{-\frac{a_b}{2} \cv^{2/3}}\,,
\ee
showing how the axion $\theta_b$ is parametrically lighter than its partner $\tau_b$.
Hence its low-energy dynamics decouples from all other fields. However $\theta_b$ cannot be used
as the inflaton \cite{kallosh} since its decay constant obtained from canonical normalisation \cite{StringAxions}:
\be
f_{\theta_b} \simeq { M_P \over \cv^{2/3} }\simeq M_{\rm KK}\,,
\label{fb}
\ee
is well below the Planck scale in the regime where one trusts effective field theory.

\section{A model with $N$ hierarchically light axions}
\label{Sec4}

\subsection{Description of the Calabi-Yau manifold}

In this section we will generalise the simplest LVS moduli stabilisation mechanism presented in section \ref{LVSstab},
obtaining a model with $N\gg 1$ hierarchically light axions which provides a microscopic realisation
of $N$-flation in IIB flux compactifications.
Following the general analysis of LVS scenarios presented in \cite{GeneralLVS},
we shall consider a CY three-fold with negative Euler number and whose volume takes the form:\footnote{See \cite{CKM} for some explicit CY examples
whose volume form looks qualitatively like (\ref{cvol}).}
\be
\label{cvol}
\cv =  \hat{\cv}(\tau_{b_i}) - \tau_s^{3/2}\,,
\ee
where $\tau_s$ is the volume of a del Pezzo divisor. On the other hand, $\hat{\cv}(\tau_{b_i})$ is a $\tau_s$-independent homogeneous function
of degree $3/2$ of $N$ four-cycles $\tau_{b_i}$, $i = 1,...,N$, which control the overall size of the CY:
\be
\hat{\vo}(\tau_{b_i}) = \frac{1}{3!}k_{ijk} t_{b_i} t_{b_j} t_{b_k}\,,
\label{GenVol}
\ee
where $k_{ijk}$ are the CY triple intersection numbers whereas $t_{b_i}$ are two-cycle volumes which are related to the four-cycle
volumes by $\tau_{b_i}=\frac 12 k_{ijk} t_{b_i} t_{b_j}$.
Similarly to section \ref{LVSstab}, we shall focus on the limit $\tau_{b_i}\gg \tau_s$ $\forall i$, so that the CY volume (\ref{cvol})
can be well approximated as $\vo \simeq \hat{\vo}(\tau_{b_i})$.
As we shall see below, each $\tau_{b_i}$ behaves qualitatively as the `big' modulus $\tau_b$ of the two moduli case described in section \ref{LVSstab}.
Hence each $\tau_{b_i}$ gets fixed due to perturbative effects, and the corresponding axion $\theta_{b_i}$ remains hierarchically lighter
with a decay constant of order $f_{b_i}\simeq M_{\rm KK}$ as in (\ref{fb}).
In this way we obtain $N$ light axions which can play the r\^ole of inflatons if $N$ is taken to be very large.
Moreover, after moduli stabilisation, the VEV of each $\tau_{b_i}$ will be of the same order of magnitude
and exponentially large in $\tau_s$:
\be
\tau_{b_i}^{3/2} \sim \tau_b^{3/2} \sim |W_0|\,e^{a_s \tau_s}\qquad\forall\,i=1,...,N\,,
\ee
and so the CY volume (\ref{cvol}) simplifies to:
\be
\vo \simeq \left(\sum_{i,j,k=1}^N k_{ijk}\right) \frac{t_b^3}{3!}\,.
\label{VolSimpl}
\ee
The cubic matrix $k_{ijk}$ has $N^3$ entries. However, in most of the known explicit CY examples built via toric geometry \cite{CKM,CMV},
only $\mc{O}(N)$ entries are non-vanishing with both positive and negative intersection numbers.
Therefore, because of possible cancellations between different non-zero entries of $k_{ijk}$, the volume form (\ref{VolSimpl}) might not
scale as $N$ \cite{nfla}. This property is crucial to be able to use the large number $N\gg 1$ to obtain slow-roll parameters smaller than unity.
In fact, by combining (\ref{eps}) with (\ref{fb}), we have:
\be
\epsilon=\eta\simeq \frac{2}{N} \left(\frac{M_P}{f}\right)^2\simeq \frac{2}{N} \,\vo^{4/3}\,,
\label{Eps}
\ee
which would behave as $N^{1/3}$ if $\vo$ scaled as $N$, showing the impossibility to get $\epsilon\simeq\eta\ll 1$ for $N\gg1$.
We shall therefore focus on CY three-folds with:
\be
\sum_{i,j,k=1}^N k_{ijk}\sim \mc{O}(1)\qquad\Rightarrow \qquad \vo\sim t_b^3 \sim \tau_b^{3/2}\,.
\ee

\subsection{Moduli stabilisation}

Let us now describe how to stabilise the K\"ahler moduli considering the case with $|W_0|$
of order unity. In this regime, perturbative corrections are naturally more important than non-perturbative
ones except for the case of the `small' modulus $\tau_s$ since a negative exponential in $\tau_s$ does not
give rise to a suppression which is strong enough to render this contribution negligible with respect to the perturbative ones.
We can therefore analyse the total scalar potential step by step, focusing first only on the dominant effects
in an expansion in inverse powers of the large volume $\vo$, and then including also subleading effects
if some flat directions are left over.\footnote{Of course we can present
our results in this fashion only a posteriori; first having analysed the complete system with all terms in the effective action included.}
The full scalar potential looks like:
\be
V_{\rm tot} = V_{\alpha'+{\rm np}_s}(\tau_s,\theta_s,\vo) + V_{g_s}(\tau_{b_i}) +V_{{\rm np}_b}(\theta_{b_i})\,,
\ee
where the leading term is generated by $T_s$-dependent non-perturbative contributions to $W$
plus $\alpha'$ corrections to $K$ and scales as:
\be
V_{\alpha'+{\rm np}_s}(\tau_s,\theta_s,\vo)\sim \mc{O}(\vo^{-3})\,.
\ee
Notice that this term depends only on $\tau_s$, $\theta_s$, and a single combination of all $\tau_{b_i}$ corresponding to $\vo$.
Hence only three moduli can be fixed at leading order, leaving $2N-1$ real flat directions out of the $2N+2$ initial ones.
The next-to-leading term is the contribution from string loop corrections to $K$ which scale as:
\be
V_{g_s}(\tau_{b_i})\sim\mc{O}(\vo^{-10/3})\,,
\ee
and are expected to depend on all $\tau_{b_i}$. Thus $N-1$ geometric moduli associated to volumes of large four-cycles get stabilised
by $g_s$ effects, leaving only $N$ axionic flat directions left over. These get lifted by tiny non-perturbative effects in each $\tau_{b_i}$
which behave as:
\be
V_{{\rm np}_b}(\theta_{b_i}) \sim \mc{O}\left(\vo^{-4/3}\,e^{-\vo^{2/3}}\right).
\ee
Hence the $N$ axions $\theta_{b_i}$ are perfect inflaton candidates since they are hierarchically lighter
than the corresponding geometric moduli $\tau_{b_i}$.

Let us now describe a bit more in detail each of these different contributions to the scalar potential.
Some material useful in performing the computations is included in appendix \ref{App}.

\subsubsection{Non-perturbative effect in $T_s$ and $\alpha'$ corrections}

The leading order contributions to the scalar potential come from
the $\alpha'$-corrected K\"ahler potential (\ref{KLVS}) and the superpotential (\ref{Lvssup})
which includes $T_s$-dependent non-perturbative effects.
After using (\ref{genpot}) and (\ref{Simiden}), the expression of the scalar potential takes the same for as (\ref{Vlvs}):
\be
V_{\alpha'+{\rm np}_s}(\tau_s,\theta_s,\vo)= \lambda_1\,\frac{\sqrt{\tau_s}\,e^{-2a_s\tau_s}}{\vo}
+\lambda_2 \frac{|W_0|\tau_s\,e^{-a_s\tau_s}}{\vo^2} \cos ( a_s \theta_s + \psi_s)+ \lambda_3\frac{|W_0|^2}{\vo^3}\,.
\label{potsmall}
\ee
Note that this potential depends on $\tau_s$, $\theta_s$ and just one combination of $\tau_{b_i}$
corresponding to $\vo\simeq\hat{\cv}(\tau_{b_i})$, while is independent of the other moduli.
Hence it stabilises only these three moduli at (see (\ref{globalMin})):
\be
\theta_s = \frac{\pi- \psi_s}{a_s}\,,\qquad \tau_s \sim \frac{1}{g_s}\,,
\qquad \vo\simeq\hat{\cv}(\tau_{b_i})\sim |W_0|\,e^{a_s/g_s}\,.
\ee
By expanding each modulus around its VEV, one can find the same mass spectrum as (\ref{ms}) and (\ref{mb}):
\be
\label{smallmass}
m_{\theta_s} \simeq m_{\tau_s}  \simeq m_{3/2}\simeq{ |W_0| \over \cv}M_P \qquad \text{and}
\qquad m_{\hat{\vo}} \simeq { |W_0| \over \cv^{3/2}}M_P\,.
\ee
Notice that $N-1$ combinations of $\tau_{b_i}$ orthogonal to $\hat{\cv}$ and the $N$
axionic fields $\theta_{b_i}$ remain exactly massless at this level of approximation.

\subsubsection{String loop effects}

Open string one-loop corrections to the K\"ahler potential have been explicitly computed only
for the case of simple toroidal orientifolds \cite{bhk} but their scaling with the K\"ahler moduli
has been generalised to arbitrary CY three-folds by exploiting their interpretation as the tree-level
propagation of closed strings \cite{bhp} and by comparing them with the behaviour of the low-energy Coleman-Weinber
potential \cite{ccq}. The $\tau_{b_i}$-dependent loop corrections to $K$ scale as:
\be
K_{g_s}\simeq \frac{1}{\vo}\sum_{i=1}^N\left(g_s c_i\, t_{b_i}+\frac{d_i}{t_{b_i}}\right),  \label{Kgs}
\ee
where $c_i$ and $d_i$ are unknown functions of the complex structure moduli, which can be considered
as constants after the $U$-moduli are stabilised at tree-level by background fluxes.
In the large volume limit $t_{b_i}\sim \vo^{1/3}$, $\forall i=1,..,N$, the first term in (\ref{Kgs})
scales as $\vo^{-2/3}$, and so dominates over the leading $\alpha'$ correction to $K$ which scales as $\vo^{-1}$,
as can be seen from (\ref{KLVS}). However, due to the `extended no-scale structure' cancellation \cite{bhp,ccq},
at the level of the scalar potential, $g_s$ effects are subleading with respect to pure $\alpha'$ corrections
which give rise to a potential of $\mc{O}(\vo^{-3}$), as can be seen from (\ref{potsmall}).
In fact, string loops generate a potential of the form:
\be
V_{g_s}(\tau_{b_i})=\sum_{i=1}^N\left[g_s^2 c_i^2 \left(\frac{t_{b_i}^2}{2\vo}-\frac{\partial t_{b_i}}{\partial \tau_{b_i}} \right)
- \frac{d_i}{t_{b_i}}\right] \frac{|W_0|^2}{\vo^3}\,,
\label{Vgs}
\ee
which, in the limit $t_{b_i}\sim \vo^{1/3}$, $\forall i=1,..,N$, scales as $\vo^{-10/3}$,
and so it is indeed subdominant with respect to the $\alpha'$ one.
We also stress that, at this level of approximation,
the extra dimensions can be considered as compact since the overall volume $\vo$ has already been stabilised at leading order.
Thus the string loop potential (\ref{Vgs}), which is expected to depend on all geometric moduli $\tau_{b_i}$,
should in general fix all of them without giving rise to any run-away direction \cite{bhp,ccq}. Moreover, if the coefficients $c_i$ and $d_i$ are $\mc{O}(1)$ constants,
the VEV of each $\tau_{b_i}$ should be of the same order of magnitude: $\tau_{b_i}\sim \tau_b$, $\forall i=1,..,N$.
By studying small fluctuations around the moduli VEVs, it is easy to realise that $N-1$ combinations of $\tau_{b_i}$ orthogonal to $\hat{\cv}$
acquire a mass of order:
\be
m_{\tau_{b_i}} \simeq { |W_0| \over \cv^{5/3}}M_P\qquad\forall i=1,...,N-1\,.
\label{loopMass}
\ee
Notice that, because of their shift symmetry, all the $N$ axionic directions $\theta_{b_i}$ are still flat after the inclusion of
string loop corrections to $K$.

\subsubsection{Non-perturbative effects in $T_{b_i}$}

The axions $\theta_{b_i}$ can acquire a mass only after the inclusion of non-perturbative effects in $T_{b_i}$
which are responsible for breaking their shift symmetries. These new tiny contributions to the superpotential read:
\be
W_{{\rm np}_b} = \sum_{i=1}^N A_{b_i}\,e^{-a_{b_i} T_{b_i}}\,,
\ee
and give rise to a subleading scalar potential of the form:
\be
\label{vaxion}
V_{{\rm np}_b} (\theta_{b_i}) = -\frac{2|W_0|}{\vo^2} \sum_{i,j=1}^N\hat{K}^{i \bar{j}}  \partial_{\bar{j}} \hat{K}a_{b_i} |A_{b_i}|
e^{-a_{b_i} \tau_{b_i}}\cos\left( a_{b_i} \theta_{b_i} + \psi_{b_i}\right),
\ee
where $\psi_{b_i}$ is defined by $A_{b_i} \overline{W}_0 =  |A_{b_i} W_0| e^{i \psi_{b_i}}$,
and $\hat{K} = -2 \ln \hat{\cv}(\tau_{b_i})$. The K\"ahler metric and its inverse (in the large volume limit) for $\hat{K}$
are given in appendix \ref{App}. Given that $\hat{\cv}(\tau_{b_i})$ is a homogeneous function of degree $3/2$, the K\"ahler potential
$\hat{K}$ obeys the identity:
\be
\label{idden}
\hat{K}^{i \bar{j}}  \partial_{\bar{j}} \hat{K} = - 2\tau_{b_i} \,.
\ee
Using this result, the potential (\ref{vaxion}) simplifies to:
\be
\label{axifinal}
V_{{\rm np}_b} (\theta_{b_i}) =  \frac{4|W_0|}{\vo^2} \sum_{i=1}^N a_{b_i}\tau_{b_i}|A_{b_i}|
e^{-a_{b_i} \tau_{b_i}}\cos\left( a_{b_i} \theta_{b_i} + \psi_{b_i}\right),
\ee
which has a minimum at $a_{b_i} \theta_{b_i} + \psi_{b_i}=\pi$, $\forall i=1,...,N$.
Using the fact that at the minimum $\tau_{b_i} \sim \cv^{2/3}$, the potential (\ref{axifinal}) clearly scales as:
\be
\label{axionapp}
V_{{\rm np}_b} (\theta_{b_i}) \sim  \mc{O}\left(\vo^{-4/3}\,e^{- a_b \cv^{2/3}}\right),
\ee
showing how for $\vo\gg 1$ this term is very suppressed with respect to all the other contributions described above.
The kinetic terms for the axions $\theta_{b_i}$ are governed by the K\"ahler potential $\hat{K}$.
In general one expects kinetic mixing with mixing matrix $\hat{K}_{i \bar{j}}$ which in the large volume limit looks like:
\be
\hat{K}_{i \bar{j}} \simeq \cv^{-4/3} H_{i \bar{j}}\,,
\label{Kscal}
\ee
where $H_{i \bar{j}}$ is a matrix with order one entries. This implies that the decay constants become:
\be
\label{axdecay}
f_{\theta_{b_i}}  \simeq {M_P \over \cv^{2/3} } \simeq M_{\rm KK}\qquad \forall i=1,...,N\,.
\ee
Combining this result with (\ref{axionapp}), one finds that the axions have mass:
\be
\label{tmass}
m_{\theta_{b_i}} \simeq  \sqrt{|W_0|} M_P\,e^{-\frac{a_{b_i}}{2} \cv^{2/3}}\qquad \forall i=1,...,N\,.
\ee
Thus at large volume the masses of the axions $\theta_{b_i}$ are significantly smaller than that
of the other moduli given in (\ref{smallmass}) and (\ref{loopMass}). This makes them ideal candidates to drive $N$-flation.
The potential for the canonically normalised axions $\hat{\theta}_{b_i}$ obtained from (\ref{axifinal}) is:
\be
\label{canpot}
V_{{\rm np}_b} (\hat{\theta}_{b_i}) =  \frac{4|W_0|}{\vo^2} \sum_{i,j=1}^N a_{b_i}\tau_{b_i}|A_{b_i}|
e^{-a_{b_i} \tau_{b_i}}\cos\left( a_{b_i} S_{ij}\hat{\theta}_{b_j} + \psi_{b_i}\right),
\ee
where $S_{ij}$ is the matrix which diagonalises the K\"ahler metric $\hat{K}_{i\bar{j}}$. Notice that because of the volume
scaling in (\ref{Kscal}), the matrix $S_{ij}$ behaves as $\vo^{2/3}$.

The part of the potential relevant for $N$-flation is the quadratic one. Expanding the canonically
normalised axions about their minima as:
\be
\label{canexp}
\hat{\theta}_{b_i} = S^{-1}_{ij} \langle\theta_{b_j}\rangle +  \varphi_i\,,
\ee
one finds:
\be
\label{quadpot}
V(\varphi_i) =  V_0 +  \frac{m_{ij}}{2} \varphi_i \varphi_j\,,
\ee
with:
\be
\label{mab}
m_{ij} = \frac{4 |W_0|}{\cv^2} \sum_{k=1}^N a_{b_k}^3 |A_{b_k}| \tau_{b_k} e^{-a_{b_k} \tau_{b_k}} S_{ki} S_{kj}
\sim\mc{O}\left(|W_0| e^{-a_b \vo^{2/3}}\right)\,,
\ee
in perfect agreement with (\ref{tmass}). The constant $V_0$ is the part of the potential independent of $\varphi_i$
which receives contributions from all terms in the potential, not just (\ref{canpot}). This in principle can be used to
tune the value of the potential at its minimum in order to obtain a de Sitter vacuum suitable
for cosmological purposes.\footnote{See \cite{dS1,dS2,dS3,dS4,dS5,dS6,dS7} for recent attempts to obtain de Sitter vacua in string theory.}

\section{Microscopic parameters and phenomenology}
\label{Sec5}

\subsection{Inflation}

Having obtained $N$ light axions, we would like to use them to drive inflation.
Our construction does not realise $N$-flation in its simplest form described in section \ref{Sec2}
since the mass matrix $m_{ij}$ (\ref{mab}) is non-trivial.
To study the dynamics of the axions, one needs to make assumptions about $m_{ij}$. Following the approach of \cite{random},
$m_{ij}$ will be taken to be a random matrix whose entries are chosen such that the axion masses (\ref{tmass}) are peaked at:
\be
\label{tmassPeak}
m \simeq  \sqrt{|W_0|} M_P\,e^{-\frac{a_b}{2} \cv^{2/3}}\,.
\ee
On the other hand, the decay constants (\ref{axdecay}) will be taken to be in a gaussian distribution peaked at $f\simeq M_{\rm KK}$.
With these assumptions, the inflationary scalar potential (\ref{quadpot}) can be evaluated in the statistical
sense \cite{random} (setting $V_0=0$):
\be
\label{InflPot}
V =  \frac 12 \langle m_{ij} \varphi_i \varphi_j \rangle \simeq \frac{m^2}{2} N f^2 \,,
\ee
obtaining an expression similar to (\ref{Vphii}) with $\rho = f_{\rm eff} = \sqrt{N} f$.
Hence the expression of the scalar potential is precisely equal to the case of $m_{ij} = \delta_{ij}$
and equal decay constants for each axion. It was found in \cite{random} that the same holds true
for the inflationary dynamics, i.e. the system can be effectively
thought of as having $N$ degrees of freedom with no cross coupling. Therefore the predictions
for the main cosmological observables are the same as those of the simplest realisation of $N$-flation \cite{nfla}
described in section \ref{Sec2}.

The requirement of obtaining around $60$ e-foldings of inflation and the right value of
the amplitude of the density perturbations, sets the value of two microscopic parameters:
the number $N=h^{1,1}-1$ of light axions and the VEV of the internal volume $\vo$.
In fact, (\ref{Nnumber}) combined with (\ref{axdecay}) gives:
\be
N\simeq 240\,\vo^{4/3}\,,
\label{Nfin}
\ee
whereas (\ref{mFix}) combined with (\ref{tmassPeak}) leads to (for $|W_0|\simeq 1$ and $a_b=2\pi/N_b$):
\be
\vo \simeq \left[\frac{2}{a_b}\ln\left(\frac{M_P}{1.6\cdot 10^{14}\,{\rm GeV}}\right)\right]^{3/2}\simeq 5.4\cdot N_b^{3/2}\,.
\ee
Choosing $N_b\simeq 10$, the CY volume turns out to be:
\be
\vo \simeq 170\,,
\ee
giving a value which is still within the regime of validity of the effective field theory which can
be trusted for $|W_0|\ll \vo^{1/3}$ \cite{border}. In fact, in our case we would have $\vo^{1/3}\simeq 5.5$
which is larger than $|W_0| \simeq 1$. Hence the Kaluza-Klein scale:
\be
M_{\rm KK}\simeq f \simeq \frac{M_P}{\vo^{2/3}} \simeq 7.8\cdot 10^{16}\,{\rm GeV}\,,
\label{fMKK}
\ee
is larger than the gravitino mass:
\be
m_{3/2} \simeq \frac{|W_0| M_P}{\vo} \simeq 1.4\cdot 10^{16}\,{\rm GeV}\,.
\ee
LVS models with volumes of order $10^2$ have been obtained in \cite{CMV}.
For this value of $\vo$, (\ref{Nfin}) gives the following number of light axions:
\be
N\simeq 2\cdot 10^5\,.
\ee
Explicit examples of CY manifolds with such a large number of K\"ahler moduli have not been constructed yet.
However, the fact that the number of topologically distinct CY three-folds is bounded or not is still an open mathematical problem.
Moreover, from the physics side, there is no strong argument suggesting that this number should be finite.

However such a large value of $N$ for $\vo \simeq 170$ gives rise to a very large density of cycles per unit volume, 
opening up the possibility to have dangerous back-reaction corrections due to a large number of species 
which are hard to estimate. This remains as the main open problem to achieve a fully trustable explicit embedding of $N$-flation in string compactifications.

\subsection{Supersymmetry breaking}

The LVS stabilisation procedure gives rise to string vacua which break supersymmetry spontaneously
along the K\"ahler moduli directions. Therefore these fields acquire non-vanishing $F$-terms and
mediate supersymmetry breaking to the visible sector via gravitational interactions.
When the visible sector lives on magnetised D7-branes wrapping internal four-cycles,
the soft terms turn out to be of order the gravitino mass \cite{CQS}:
\be
M_{\rm soft}\simeq m_{3/2} \simeq 1.4\cdot 10^{16}\,{\rm GeV}\,,
\ee
whereas when it is localised on fractional D3-branes at singularities,
the soft terms can be very suppressed with respect to the gravitino mass due to sequestering effects \cite{sequestering}:
\be
M_{\rm soft} \simeq m_{3/2}\left(\frac{m_{3/2}}{|W_0| M_P}\right) \simeq \frac{|W_0| M_P}{\vo^2}\simeq 8.4\cdot 10^{13}\,{\rm GeV}\,.
\ee
In both cases, the soft terms are at a very high scale, showing the presence of a strong tension between $N$-flation
and TeV-scale supersymmetry. This is a generic feature of models with large tensor-to-scalar ratio where the inflationary
scale is just slightly below the GUT scale. Notice that even sequestering does not help to alleviate this tension considerably,
implying that $N$-flation predicts no discovery of superpartners at the LHC and, more deeply,
a solution to the hierarchy problem that is not based on low-energy supersymmetry.

\subsection{Reheating}

After the end of inflation, the Standard Model degrees of freedom get excited during the reheating
process by the decay of the lightest scalars. In our model, these are
the $N$ axions which drive inflation. If the visible sector is built via D7-branes
wrapped on the large four-cycles whose volumes are given by the geometric moduli $\tau_{b_i}$,
the canonically normalised axions $\varphi_i$ couple to gauge bosons via the topological term \cite{StringAxions}:
\be
\mc{L}_{\varphi_i\gamma_j\gamma_j}=\frac{\alpha_j}{8\pi}\frac{\varphi_i}{f_i}F_j \wedge F_j\,,
\label{phigg}
\ee
where the gauge couplings are given by $\alpha_j^{-1}=4\pi g_j^{-2}=\tau_j$.
Moreover, the axions interact with matter fermions with derivative
couplings of the form \cite{StringAxions}:
\be
\mc{L}_{\varphi_i\psi_j\psi_j} = \frac{\partial_{\mu} \varphi_i}{2 f_i} \bar{\psi}_j \gamma^{\mu} \gamma_5  \psi_j\,.
\label{phipsipsi}
\ee
The coupling to gauge bosons (\ref{phigg}) gives rise to a decay width of order:
\be
\Gamma_{\varphi_i\to\gamma\gamma}\simeq N_\gamma \frac{m^3}{f^2}\qquad\forall i=1,...,N\,,
\label{decay}
\ee
where $N_{\gamma}=12$ is the number of Standard Model gauge bosons.
On the other hand, the decay rate to matter fermions from (\ref{phipsipsi}) scales as:
\be
\Gamma_{\varphi_i\to\psi\psi}\simeq \frac{m m_\psi^2}{f^2}\qquad\forall i=1,...,N\,,
\ee
which for $m_\psi\ll m$ is negligible with respect to the decay to gauge bosons (\ref{decay}).
The $N$ light axions decay simultaneously when $ H \simeq \Gamma_{\varphi_i\to\gamma\gamma}$,
dumping to the visible sector an energy density of order:
\be
\rho_{\rm rad} \simeq H^2 M_P^2 \simeq \Gamma_{\varphi_i\to\gamma\gamma}^2 M_P^2\,.
\ee
Hence the visible sector is reheated to a temperature:
\be
T_{\rm rh}\simeq \rho_{\rm rad}^{1/4}
\simeq \frac{m}{f} \sqrt{N_\gamma m M_P}\simeq 1.4\cdot 10^{14}\,{\rm GeV}\,,
\ee
which turns out to be very high. Even though this temperature is very high compared to the usual reheating temperature, there is no associated gravitino overproduction problem as the associated gravitini are heavier than this temperature. Notice that in general the $N$ light axions could also
decay to hidden sector degrees of freedom, producing dark radiation \cite{Cicoli:2012aq,Higaki}.
The main candidate for dark radiation in our model is the QCD axion which could be one of the $N$ light axions,
say $\varphi_{\rm QCD}$, if the mass developed by $T_{b_{\rm QCD}}$-dependent non-perturbative effects
is smaller than the mass generated by standard QCD instantons:
\be
m_{\rm QCD} \simeq \frac{\Lambda_{\rm QCD}^2}{f} \simeq 0.5\,{\rm neV}\,.
\ee
Notice that this mass is very small because the axion decay constant (\ref{fMKK}) is slightly larger than the GUT scale.
In order to avoid axionic dark matter overproduction for such a large $f$, one has to allow for some tuning
of the the initial misalignment angle of the QCD axion. It can be easily seen that (\ref{tmass}) can be smaller than $m_{\rm QCD}$
if $|W_0|=1$ and $a_{b_{\rm QCD}}=2\pi$ since one would obtain $m_{\varphi_{\rm QCD}}\simeq 4\cdot 10^{-6}\,{\rm neV}$.

\subsection{Quintessence}

At the present epoch, the energy density of the universe is dominated by dark energy.
The most popular proposal for a microscopic description of dark energy is a cosmological constant term in the Lagrangian.
An alternative description is provided by models of quintessence where the late time cosmology of the universe
is governed by a scalar field slowly rolling to its minimum (usually taken to be at zero energy density).
Given that the dynamics of quintessence is very similar to that of inflation, the two cosmological scenarios have the same kind of UV sensitivity. The idea of quintessence using a single axionic field was introduced in \cite{Frieman:1995pm}, but similarly to inflation, observations of the Hubble scale using type Ia supernovae data push the axion decay constant $f$ to be close to $M_P$ \cite{Dutta:2006cf}. Note that the constraint for quintessence is not as severe as that for inflation where $f$ needs to be super-Planckian to be in accordance with observations. A detailed study of quintessence phenomenology with large number of pseudo Nambu-Goldstone bosons, along the line of $N$-flation, can be found in \cite{kale}.

Our construction provides also a natural setting for quintessence, in particular $N$-quintessence. Note that by combining the expression for the slow-roll parameters (\ref{eps}) with the one for the Hubble
constant (\ref{H}), one can write:
\be
\label{hubvol}
3\eta H_{\rm quint}^2 \simeq m^2\,.
\ee
Demanding the present Hubble constant to be of order $H_{\rm quint}\simeq 10^{-60} M_P$ and taking $\eta \simeq 10^{-x}$ with $x>0$, one finds:
\be
\label{qcri}
m \simeq 10^{-\left(60+\frac{x}{2}\right)} M_P\,.
\ee
This gives the following constraints for the two microscopic parameters $\vo$ and $N$ (for $a_b=2\pi/N_b$):
\be
\vo \simeq  3\left(1 + \frac{x}{80}\right) N_b^{3/2}\cdot 10^2\qquad\text{and}\qquad
N \simeq 4\left(1 + \frac{x}{60}\right) N_b^2\cdot 10^{3 + x}\,.
\ee
For $x\simeq 1$ and $N_b\simeq 10$ one obtains $m\simeq 10^{-30}\,{\rm meV}$, $\vo\simeq 4\cdot 10^3$ and $N\simeq 1\cdot 10^6$.
We clearly see that just by varying the overall volume $\vo$ from values of $\mc{O}(10^2)$ to values of $\mc{O}(10^3)$,
the energy scale of the axion potential (\ref{vaxion}) decreases dramatically because of the exponential suppression in $\vo^{2/3}$.
Hence we can easily use our set-up to realise models of both inflation and quintessence.
We finally mention that our construction is not incompatible with bounds from fifth forces
even if the axions are extremely light with masses of order $10^{-30}\,{\rm meV}$.
In fact, given that the axion is a pseudo scalar, the interaction (\ref{phipsipsi}) is spin dependent,
and so the force between macroscopic unpolarised objects (such as the earth or the sun) vanishes.
Therefore fifth-force bounds do not impose any strong constraint.

\section{Conclusions}

Being a pseudo-Nambu Goldstone boson whose mass is protected by an approximate shift symmetry,
the axion is the most well motivated inflaton candidate. The stability of the inflaton mass
due to quantum corrections is under control because of the shift symmetry. However, recent CMB observations push the axionic decay constant $f$ to be super-Planckian. On the other hand, there is a theoretical bound on $f$ being less than $M_P$.
In this situation, the rescue comes from the idea of $N$-flation, where one uses a large number of axions which collectively drive inflation.
The effective decay constant of the collective degree of freedom becomes super-Planckian, satisfying the observations, but individual decay constants $f_i < M_P$. At the same time, even though each field traverses $\Delta \phi_i < M_P$, the effective field moves through
a super-Planckian distance, making it observationally promising by producing large tensor perturbations.

In this paper, we have proposed a model of $N$-flation in the context of the LARGE Volume Scenario of type IIB compactifications. The most important issue about the construction of $N$-flation is the r\^ole of the geometric moduli which come in pair with axionic fields.
In earlier attempts of constructing $N$-flation in string theoretic set-ups, the issue of stabilising these moduli
has not been addressed in a satisfactory manner \cite{kallosh}.
The issue of moduli stabilisation is more pressing when the scale of inflation is high as in $N$-flation since
all the moduli other than the inflaton must be stabilised at a much higher scale.
In our construction all geometric moduli receive masses which are parametrically larger than the axion masses
and the Hubble scale during inflation. This effectively decouples the geometric moduli
from the inflationary dynamics at its onset.

As we have noted earlier, in KKLT-like constructions, both axions and geometric moduli receive masses of the same order,
making this scenario unsuitable for a simple realisation of $N$-inflation. In the LVS case with two moduli and CY volume of the Swiss cheese form,
a large hierarchy between the mass of the axions and the geometric moduli is realised naturally,
even if the axionic decay constant is below the Planck scale.

In our construction, we generalised the simplest LVS scenario obtaining $N$ hierarchically light axions suitable to drive $N$-flation.
The complex structure moduli and the dilaton are fixed by background fluxes and acquire masses at a very high scale compared to the inflationary scale.
Our model involves $N$ K\"ahler moduli $T_{b_i}=\tau_{b_i}+{\rm i}\theta_{b_i}$ associated to large four-cycles of the CY,
and the modulus $T_s=\tau_s+{\rm i}\theta_s$ related to a small blow-up mode. Non-perturbative effects associated to the small cycle
stabilise both $\tau_s$ and $\theta_s$. On the other hand, the interplay between these non-perturbative effects
and $\alpha'$ corrections to the K\"ahler potential fixes a combination of geometric moduli that corresponds to the total CY volume.
All the other $(N-1)$ fields $\tau_{b_i}$ are fixed by string loop effects at subleading order at approximately
the same value: $\tau_{b_i}\simeq\vo^{2/3}$ \cite{GeneralLVS}.
The important point here is that these loop effects respect the axionic shift symmetry, and so at this stage,
even though all geometric moduli are stabilised, $N$ axionic directions still remain flat.
These fields develop their potentials from tiny non-perturbative effects which depend on the `big' moduli $T_{b_i}$.
We use these $N$ slow-rolling light degrees of freedom with sub-Planckian decay constants for our model of $N$-flation.

Phenomenological considerations require the volume to be $\vo \sim \mc{O}(10^2)$
in string units with a relatively large number of axions $N \sim \mc{O}(10^5)$.
In this way the Hubble scale during inflation turns out to be very large: $H_{\rm inf}\sim \mc{O}(10^{15})$ GeV
corresponding to a GUT-scale gravitino mass: $m_{3/2}\sim \mc{O}(10^{16})$ GeV. This implies that the
soft terms generated by gravity mediation are also very large, and so undetectable by the LHC, since they would
be of order the gravitino mass. Even a possible sequestering of the visible sector from supersymmetry breaking
would not help to obtain TeV-scale supersymmetry since the soft mass would be lowered only to energies of $\mc{O}(10^{14})$ GeV.
We therefore realised that high-energy supersymmetry is a typical feature of axionic $N$-flation.

The major problem to be solved in order to realise a fully explicit embedding of $N$-flation in type IIB string compactifications 
following our moduli stabilisation mechanisms which creates a mass hierarchy between axions and volume moduli, 
is the computation of all possible back-reaction issues due to a large number of species.

From the observational point of view, $N$-flation is very promising since it produces large tensor modes in agreement with recent BICEP2 data \cite{BICEP2}.
A striking prediction of $N$-flation is that the tensor-to-scalar ratio $r \simeq 8/N_e$ is independent of the number of fields,
their masses and initial conditions. $N$-flation thus makes a definite predictions for $r \sim 0.13$ assuming $N_e \simeq 60$,
and most importantly, the predicted value is well within the reach of observations \cite{Kim:2006ys,Alabidi:2005qi}.
It is interesting, although not surprising, to note that this value is similar to the chaotic inflationary scenario with quadratic potential.
This is simply due to the fact that the effective degree of freedom behaves like a quadratic potential with super-Planckian field excursion
since the effect of cross-couplings is negligible \cite{random}. On the other hand, the prediction for the spectral index is dependent on model parameters and initial conditions. Typically the spectral index for $N$-flation is more red compared to the single field chaotic inflation prediction $n_s \simeq 0.96$, but this is usually well within recent observational limits \cite{Kim:2006ys}. With the quadratic approximation of the full potential, the predicted non-gaussianities are unobservably small. Recently it has been noted that the full cosine potential can have dramatic effects on observable quantities, notably reducing the value of the tensor-to-scalar ratio, but making non-gaussianities observably large \cite{Kim:2010ud,Kim:2011jea}.

An important aspect of $N$-flation is its reheating mechanism. It turns out that preheating via parametric resonance is ineffective for the case of $N$-flation. This is mainly due to incoherent contributions from different axion fields in the production of matter particles \cite{Battefeld:2008bu}. In this paper we outlined the effect of perturbative reheating showing that one can have a very efficient reheating of the visible sector degrees of freedom
produced by the decays of the $N$ light axions which drive inflation. The leading couplings originate from dimension five operators suppressed by the axionic decay constants. Finally we showed that our set-up can describe also the late-time evolution of our universe, producing a successful quintessence
model just by taking the two microscopic parameters $\vo$ and $N$ one order of magnitude larger than in the inflationary case.

Summarising our work, we have presented a set-up of axionic $N$-flation and $N$-quintessence in type IIB LVS compactifications where perturbative effects stabilise the geometric moduli which therefore turn out to be hierarchically heavier than the corresponding $N\gg 1$ axions whose collective dynamics drives inflation.

\section*{Acknowledgements}

We  would like to thank Daniel Baumann, Sven Krippendorf, Sudhakar Panda, Fernando Quevedo and Roberto Valandro for discussions. AM would like to thank the University of Bologna and the ASICTP for hospitality. KD is partially supported by a Ramanujan Fellowship and Max Planck Society-DST Visiting Fellowship. AM is partially supported supported by a Ramanujan Fellowship.

\appendix

\section{Appendix}
\label{App}

In this appendix we give some material useful for performing the computations
of section \ref{Sec4}. In LVS, the superpotential and K\"ahler potential
have the form:
\be
W =  W_0 +  \sum_I  A_I e^{-a_I  T_I},  \qquad K = - 2 \ln \left(  \cv(\tau_I) + \frac{\hat{\xi}}{2} \right)\,,
\ee
where $\hat{\xi}\equiv \xi\,g_s^{-3/2}$. With this the scalar potential can be written as:
\bea
V &=& e^K \bigg[K^{T_J \bar T_K}\left( a_J A_J a_K \bar A_K e^{-(a_J T_J + a_K \bar T_K)}
- \left(a_J A_J e^{-a_J T_J} \bar W \partial _{\bar T_K}  K+ a_K \bar A_K e^{-a_K \bar T_K} W \partial_{T_J} K\right) \right) \cr
& & + \frac{3 \hat{\xi} (\hat{\xi}^2 + 7 \hat{\xi} + \vo^2)}{(\vo - \hat{\xi})(2 \vo + \hat{\xi})^2} |W|^2 \bigg]
  \equiv V_{\rm{np}1} + V_{\rm{np}2} + V_{\alpha'}.
\label{genpot}
\eea
We also record the tree-level K\"ahler metric and its inverse (in the large volume limit)
for CY three-folds with a blow-up cycle corresponding to the resolution of a point like singularity.
As in section \ref{Sec4}, $\tau_s$ will denote the volume of the blow-up cycle and
$\tau_{b_i}$, $i = 1,...,N$, the volumes of the other cycles:
\be
\label{kaltree}
K_{\rm tree} = - 2 \ln \cv = -2 \ln \left( \hat{ \cv}(\tau_{b_i})  - \tau_s^{3/2} \right).
\ee
In the large volume limit:
\be
\label{kapp}
K_{\rm tree} \simeq \hat{K} +  2   {\tau_s^{3/2} \over \hat{\vo}},
\ee
where we have defined:
\be
\hat{K} = - 2 \ln \hat{\vo}.
\ee
It will be useful to think of $\hat{K}$ as defining a K\"ahler potential for the fields $T_{b_i}$.
The K\"ahler metric (to leading order in $\delta \simeq{\tau_s^{1/2} / {\hat{\vo}}^{1/3}}\ll 1$)
reads:
$$
{K}_{I \bar{J} } = \begin{pmatrix} \hat{K}_{i \bar{j} } &  { 3 \tau_s^{1/2}  \over {4 {\hat{\vo}}} } \hat{K}_i \\
{ 3 \tau_s^{1/2}   \over {4 {\hat{\vo}}}} \hat{K}_{\bar{j}}  & { 3 \over {8  {\hat{{\cal{V}}}} \tau_s^{1/2}} } \end{pmatrix}
\approx { 1 \over \hat{{\vo}}^{4/3} } \begin{pmatrix} {\mc{O}} ( \delta^{0})  &{\mc{O}} ( \delta^{1}) \\
{\mc{O}} ( \delta^{1})  & {\mc{O}} ( \delta^{-1}) \end{pmatrix}.
$$
In the above, capital indices run over all the moduli $(T_{b_i}, T_s)$, while lower-case ones run
only over the `big' moduli $T_{b_i}$. The inverse K\"ahler metric is:
$$
{K}^{I\bar{J} } = \begin{pmatrix} \hat{K}^{i \bar{j} } &  -  2 \tau_s \hat{K}^{i \bar{k}}  \hat{K}_{\bar{k}} \\
-  2  \tau_s \hat{K}^{\bar{j} {k}} \hat{K}_k    & { {8 {\hat{\vo}} \tau_s^{1/2}}  \over  3} \end{pmatrix}
\approx \hat{\vo}^{4/3} \begin{pmatrix} {\mc{O}} ( \delta^0)  &{\mc{O}} ( \delta^2) \\
{\mc{O}} ( \delta^2)  & {\mc{O}} ( \delta^1)
\end{pmatrix}.
$$
Finally, note that since $\hat{\vo}$ is a homogeneous function of $\tau_{b_i}$,
$\hat{K}$ satisfies a no-scale condition with respect to the large moduli:
\be
\hat{K}^{\bar{i}j} \hat{K}_{\bar{i}} \hat{K}_j = 3\,,
\ee
which can be used to show the following identity (to leading order in $\delta$):
\be
{K}^{i \bar{s}} K_i =   -  2 \tau_s \hat{K}^{i \bar{j}} \hat{K}_{\bar{j}}  \hat{K}_i = - 6 \tau_s\,.
\label{Simiden}
\ee

\end{document}